\documentclass[oneside,reqno,a4paper,12pt]{article}
\usepackage{latexsym}
\usepackage{euscript}
\usepackage{epsfig}
\date{\today}
 \large\normalsize

\def\ie{{\it i.e.}}

\def\npb#1#2#3{    {\it Nucl. Phys. }{\bf B #1} (19#2) #3}

\def\plb#1#2#3{    {\it Phys. Lett. }{\bf B #1} (19#2) #3}
\def\prd#1#2#3{    {\it Phys. Rev. }{\bf D #1} (19#2) #3}
\def\prep#1#2#3{   {\it Phys. Rep. }{\bf #1} (19#2) #3}
\def\prl#1#2#3{    {\it Phys. Rev. Lett. }{\bf #1} (19#2) #3}

\begin{document}
\begin{flushright}
hep-ph/9910408\\
\end{flushright}
\vskip 1.0cm
\begin{center}
{\large \bf Non-Universal Gaugino Phases and the LSP
Relic Density}
\vskip 1.5cm
{  Shaaban Khalil }\\
\vspace*{0.8cm} \small{\textit{Departmento de Fisica Te\'orica, C.XI,
Universidad Aut\'onoma de Madrid,\\ 28049 Cantoblanco, Madrid, Spain.}} \\
\vspace*{2mm}
and\\
\small{\textit{Ain Shams University, Faculty of Science, Cairo
11566, Egypt.}} \\

%\end{center}
\vspace*{2cm}
%\begin{center}
%{\bf Abstract}
%\end{center}
\vspace*{8mm}
%\begin{quotation} 
\begin{abstract}
{
The cosmological relic density of the lightest supersymmetric particle  
(LSP) of type I string derived model is calculated. This model can
accommodate large values of CP violating phases, and the electron and
neutron electric dipole moments satisfy the experimental constraint. We
show that the constraint from the electric dipole moment on the ratio
between the gaugino masses implies that the mass of the LSP, which is
bino like, is close to the lightest chargino. The co-annihilation between
them is very important to reduce the LSP relic density to an interesting
region. We show that the SUSY phases, although they are large, have no
significant effect on the relic density and on the the upper bound imposed
on the LSP mass. However, they are very significant for the detection
rates. We emphasize that the phase of the trilinear coupling increase the
direct and indirect detection rates.}   
\end{abstract}

%\end{quotation}

%\setcounter{page}{1}
%\end{titlepage}
\end{center}

\thispagestyle{empty}

\newpage

\section{{\bf Introduction}}
It is well known that supersymmetric (SUSY) models allow for new
possibilities for CP violation. The soft SUSY breaking terms are in
general complex. For instance, in the minimal supersymmetric model (MSSM),
there are complex phases in the parameters $A$, $B$ (which are the
coefficients of the SUSY breaking of the trilinear and bilinear couplings
respectively), $M$ (the gaugino masses), and $\mu$ (the mass coefficient
of the bilinear terms involving the two Higgs doublets). However, only two
of these phases are physical (they can not be rotated away
by all possible field redefinitions). These phases give large one-loop
contributions to the electric dipole moments (EDM) of the
neutron and electron which exceed the current limits 
\begin{eqnarray}
d_n &<& 6.3 \times 10^{-26} \mathrm{e cm}, \nonumber\\
d_e &<& 4.3 \times 10^{-27} \mathrm{ e cm}.
\label{limit}
\end{eqnarray} 
Hence SUSY phases, which are generally quite constrained by~(\ref{limit}),
have to be of order $10^{-3}$ for SUSY particle masses of order the weak
scale~\cite{dugan}. 
\vskip 0.25cm
However it was pointed out that there are internal cancellations among
various contribution to the EDM (including the chromoelectric and purely
gluonic operator contributions) whereby allowing for large
phases~\cite{nath}. We have shown that in the
effective supergravity (derived from string theory) such cancellations are
accidental and it
only occurs at few points in the parameter space~\cite{barr}. Recently, it was argued that the
non-universal gaugino masses and their relative phases are crucial for
having sufficient cancellations among the contributions to
EDMs~\cite{kane}. These cancellations have
been studied in the framework of a D-brane model where $SU(3)_C
\times U(1)$ and $SU(2)$ arise
from one five brane sector and from another set of five
branes respectively ~\cite{kane,arnowitt}. This model leads to non
universal gaugino mass which is
necessary to ensure these cancellations. 
\vskip 0.25cm
In such a case one expects that these large phases have important
impact on the lightest supersymmetric particle (LSP) relic density and
its detection rates.
In Ref.~\cite{shafi,falk} the effect of SUSY phases on the LSP mass,
purity, relic density,
elastic cross section, and detection rates has been considered within
models with universal,
hence real, gaugino masses. It was shown that the phases have no
significant effect on the
LSP relic abundance but, however, they have a substantial impact on the 
LSP detection rates.     
\vskip 0.25cm 
In this paper we study the cosmological implications of the gaugino
phases. In particular, we consider the D-brane model,
recently proposed in Ref.~\cite{munoz}, which allows large value
of SUSY phases without exceeding the experimental upper limit on the 
the neutron and electron EDMs~ Ref.~\cite{kane, arnowitt}. It turns out
that the LSP of this model, depending on the ratio between $M_1$ and
$M_2$, could be bino or wino like. In the region where the EDMs satisfy
the upper (\ref{limit}), the mass of the LSP is very close to the
lightest chargino. Hence, in this case, the co-annihilation between the LSP
and the lightest chargino becomes very important and it
greatly reduces the relic
density. The phases have no important effect on the LSP relic abundance
as in the case studied in 
Ref.~\cite{falk}. However, their effect on the detection rates is very
significant and it is     
larger than what is found in the case of real gaugino masses~\cite{falk}.
\vskip 0.25cm
In section 2 we briefly review the formula of the soft SUSY breaking
terms in $D$-brane model~\cite{munoz}. We also study the effect of the
phases on the LSP mass and its 
composition. In section 3 we compute the relic density of the LSP
including the co-annihilation
with the lightest chargino. We also comment on the effect of SUSY
phases on the LSP detection rate. We give our conclusions in section 4. 
\section{{\bf Non-universal gaugino masses}}
In the framework of the MSSM and the minimal supergravity (SUGRA) the universality of gaugino
masses is usually assumed, \ie, $M_a=M_{1/2}$ for $a=1,2,3$. Despite the
simplicity of this 
assumption, it is a very particular case and there exist classes of model
in which non-universal
gaugino masses can be derived~\cite{randall}.
\vskip 0.25cm
A type I string derived model, which has recently proposed in 
Ref.~\cite{munoz} leads to non-universal gaugino masses. This property, as
emphazised in Ref.~\cite{kane, arnowitt}, is very important
for the cancellation mechanism, mentioned in the previous section.  It has
been shown that for this 
model there exists  a special region in the parameter space
where both the electron and the neutron EDMs
satisfy the experimental constraint (\ref{limit}) and large values of
SUSY phases and light SUSY
spectrum are allowed.
\vskip 0.25cm
The soft SUSY breaking terms in type I string theories depend on the
embedding of the standard
model (SM) gauge group in the D-brane sector. In case of the SM gauge
group is not associated
with a single set of branes the gaugino masses are
non-universal~\cite{munoz}. If the $SU(3)_C$
and $U(1)_Y$ are associated with one set of five branes (say $5_1$) and
$SU(2)_L$ is associated with a second set $5_2$ . The soft SUSY breaking
terms take the following form~\cite{munoz}
\begin{eqnarray}
M_1 &= & \sqrt{3} m_{3/2} \cos \theta \Theta_1 e^{-i \alpha_1} =
M_3 = - A ,\\ M_2 &=& \sqrt{3} m_{3/2} \cos \theta \Theta_2 e^{-i
\alpha_2} ,
\end{eqnarray}
where $A$ is the trilinear coupling. The soft scalar masses squared
are given by
\begin{eqnarray}
m_Q^2 &=& m^2_L = m_{H_1}^2 = m_{H_2}^2 = m_{3/2} (1-3/2 \sin^2
\theta) ,\\ m_D^2 &=& m^2_U = m_E^2 = m_{3/2} (1-3 \cos^2 \theta),
\end{eqnarray}
and $\Theta_1^2 + \Theta_2^2 = 1$. In this case, by using the
appropriate field redefinitions and the $R$-rotation we end up
with four physical phases, which can not be rotated away. These
phases can be chosen to be: the phase of $M_1$ ($\phi_1$), the
phase of $M_3$ ($\phi_3$), the phase of $A$ ($\phi_A$), and the
phase of $\mu$ ($\phi_{\mu}$). The phase of $B$ is fixed by the
condition that $B \mu$ is real. We notice that at the GUT scale
$\phi_1=\phi_3=\phi_A = \alpha_1 -\alpha_2$ while the phase of
$\mu$ is arbitrary and scale independent.
\vskip 0.25cm  
The effect of these phases on the EDM of the electron and the neutron
(by taking into account the cancellation mechanism between the
different contributions), has been examined in Ref.~\cite{kane,arnowitt}. 
It was shown that large values of these phases can be accommodated and 
the electron and neutron EDMs satisfy the experimental constraint. It is 
interesting to note, however, that the EDMs impose a constraint on the
ratio $M_1/M_2$. In fact, in order to have an overlap between the electron
and neutron EDM allowed regions, 
$M_2$ should be smaller than $M_1$. In particular, as explained in
Ref.~\cite{kane}, a precise overlap between
these two regions occurs at $\Theta_1 = 0.85$. Such constraint has an
important impact on the LSP. In this case we have the following ratios of the
gaugino masses at the string scale
\begin{equation}
\vert M_3 \vert :\vert M_2 \vert : \vert M_1 \vert = 1:
\frac{\Theta_2}{\Theta_1} : 1 ,
\end{equation}
where $\frac{\Theta_2}{\Theta_1} <1$. So that $M_2$ is the
lightest gaugino at GUT scale. However, at the weak scale we
approximately have
\begin{equation}
\vert M_3 \vert :\vert M_2 \vert : \vert M_1 \vert = 7: 2
\frac{\Theta_2}{\Theta_1} : 1 ,
\end{equation}
since $\alpha_1 : \alpha_2 :\alpha_3 \simeq 1:2:7$ at $M_Z$. In
Figure (1) we show the running values for $\vert M_i \vert$ with
$m_{3/2}$ of order 100 GeV and $\Theta_1=0.85$. In our analysis we
restrict ourselves to the region found in Ref.~\cite{kane}, where the
electron and neutron EDMs are smaller than the
limit (\ref{limit}), \ie, we take $\tan \beta \simeq 2$, $\theta=0.2$,  
$\Theta_1 = 0.85$, $\phi_{\mu} \simeq 10^{-1}$ and $\phi_1 \simeq (1-1.5 
\pi)$.

\begin{figure}[h]
\psfig{figure=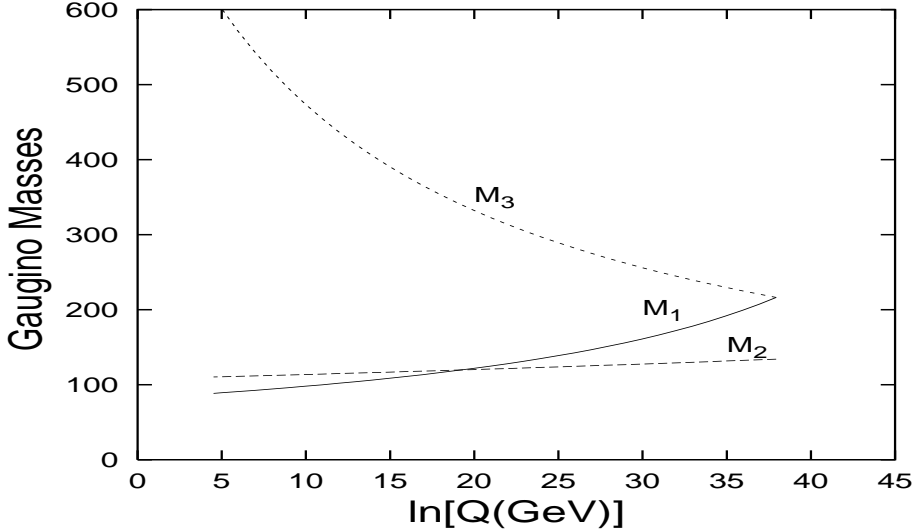,height=7cm,width=12cm} \caption{The running
values for $\vert M_i \vert $ from GUT scale to the weak scale}
\end{figure}
\vskip 0.3cm This restriction suggests that in this scenario the lightest
neutralino is a bino like. Indeed we find that the lightest neutralino,
which in general is a linear combination
of the Higgsinos $\tilde{H}_1^0$, $\tilde{H}_2^0$ and the two neutral
gaugino $\tilde{B}^0$ (bino) and $\tilde{W}_3^0$ (wino)
$$\chi = N_{11}\tilde{B}+ N_{12}\tilde{W}^3+
N_{13}\tilde{H}_1^0 + N_{14}\tilde{H}_2^0 , $$
is bino like with the gauge function $f_g= \vert N_{11}\vert^2 + \vert
N_{12} \vert^2 \simeq 0.98$.
Moreover it turns out that the LSP mass is close to the lightest chargino
mass which is equal to the mass of the next lightest neutralino
$(\tilde{\chi}_2^0)$. Figure (2) shows that the mass splitting between LSP
and the lightest chargino $\Delta m_{\chi^+} = m_{\chi_1}^+/m_{\chi} -1$ 
is less than $ 20 \%$. Therefore the co-annihilations between the bino and
the chargino, as well as the next to lightest neutralino, are very
important and have to be included in the calculation of the relic density.
\begin{figure}[h]
\psfig{figure=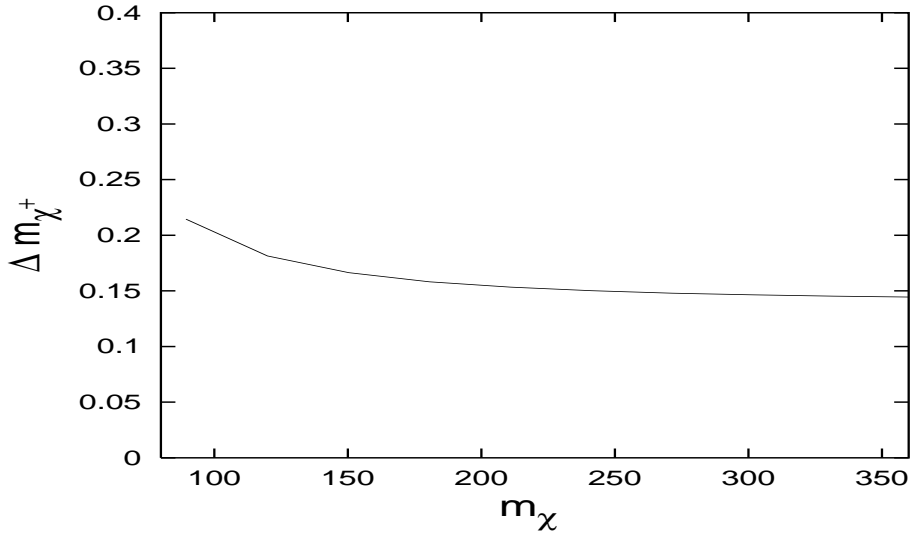,height=7cm,width=12cm}
\caption{The mass splitting $\Delta m_{\chi^+} = m_{\chi_1}^+/m_{\chi} -1$
versus the LSP mass.}
\end{figure}
\section{{\bf Relic Abundance and Co-annihilation effect}}
In this section we compute the relic density of the LSP. Moreover, we
study the effect of the SUSY CP violating phases and the co-annihilation
on both the relic density and the upper bound of the LSP mass. As
usual, since the LSP is bino like, the annihilation is predominantly  into
leptons by the exchange of the right slepton. Without the co-annihilation,
the constraint on the relic density $0.025 < \Omega_{LSP} h^2 < 0.22 $
imposes sever constraint on the LSP mass, namely
$m_{\chi} < 150$ GeV. As shown in Figure (3), the SUSY phases have
no any significant effect in relaxing such sever constraint. 
\begin{figure}[h]
\psfig{figure=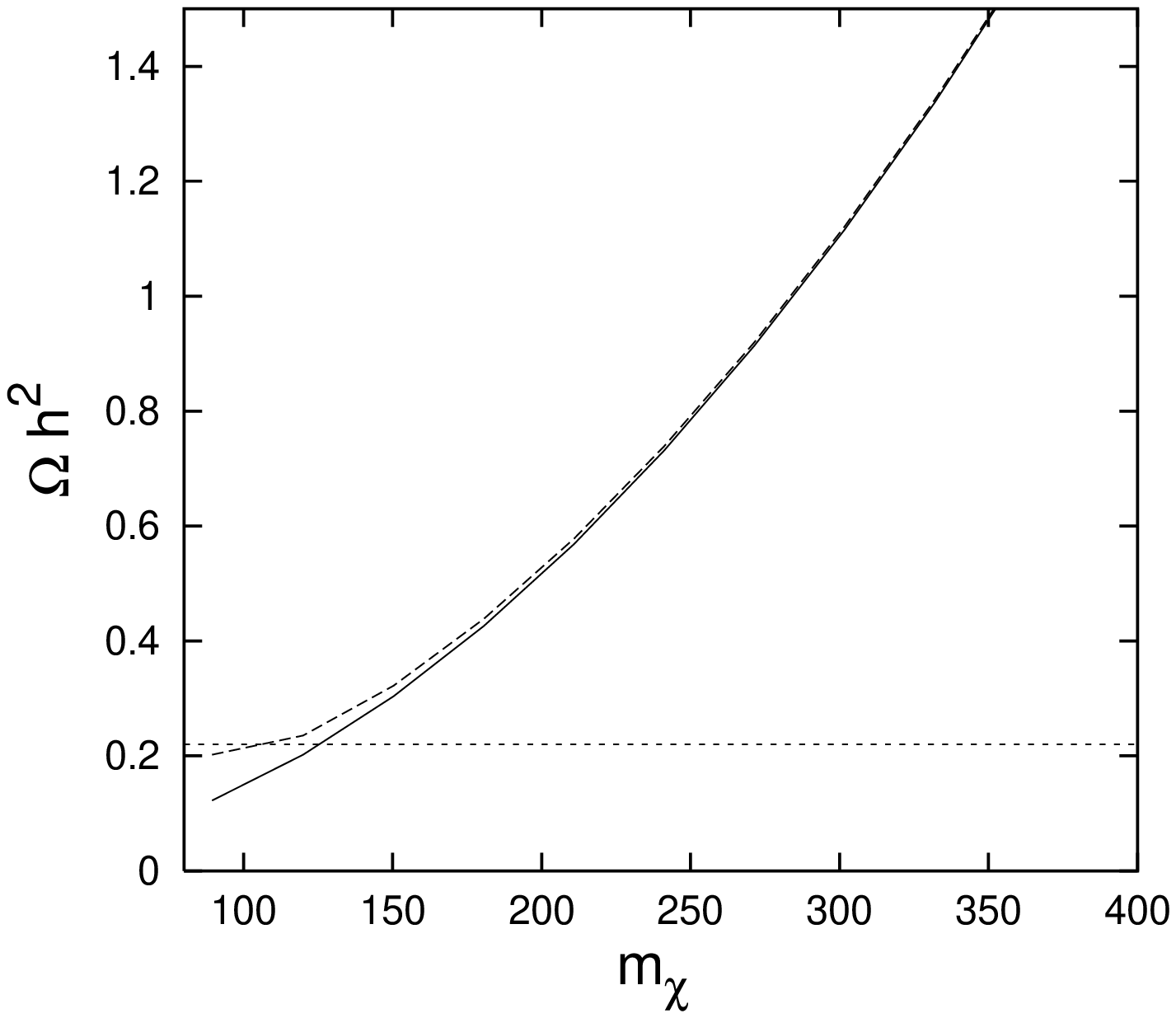,height=7cm,width=12cm}
\caption{The LSP relic abundance without co-annihilation versus is mass, solid line corresponds 
to non vanishing phases while the dashed lines correspond to vanishing phases. }
\end{figure}

We now turn to the calculation of the cosmological relic density of the
LSP by including the co-annihilation of $\chi$ with $\chi_1^+$ and
$\tilde{\chi}_2^0$. As shown in Figure (3) the  $\Omega_{LSP} h^2$
increases to unacceptable high values as $m_{\chi}$
approaches $300$ GeV. This results imposes severe constraints on the 
entire parameter space. Therefore, in order to reduce the LSP
relic density to an acceptable level, it is very important to include the 
co-annihilation. Several studies, which explain the effect of the
co-annihilation with the next to lightest SUSY particle (NLSP), have been
recently reported~\cite{george}. In these studies it was shown that, in
the models with large $\tan\beta$, the NLSP turns out to be stau and its
co-annihilation with the LSP is crucial to reduce the relic density to an
acceptable region. 
\vskip 0.25cm
By following Ref.~\cite{grist}, we define the effective number of the LSP
degrees of freedom
\begin{equation}
g_{eff} = \sum_i g_i (1+\Delta_i)^{3/2} e^{-\Delta_i x},
\end{equation}
where $\Delta_i = m_i/m_{\tilde{\chi}_1^0} -1 $, $x =
m_{\tilde{\chi}_1^0}/T$ with $T$ is the
photon temperature and $g_i=2,4,2 (i=\tilde{\chi}_1^0, \tilde{\chi}_1^+,
\tilde{\chi}_2^0)$ is
the number degrees of freedom of the particles. Note that the
neutralinos $\chi_{1,2}^0$ and chargino $\chi^{\pm}$, which are Majorana
and Dirac fermions, have two and four degrees of freedom
respectively. The Boltzmann equation
for the
total number density $n= \sum_i n_i$ is given by 
\begin{equation}
\frac{d n}{d t} = -3 H n - \langle \sigma_{eff} v \rangle (n^2 -(n^{eq})^2),
\end{equation}
where $H$ is the Hubble parameter, $v$ is the relative velocity of the
annihilation particles. The number density in the thermal equilibrium
$n^{eq}$ is given by $ n_i/n \simeq n_i^{eq}/n^{eq} = r_i $.  The
effective cross section, $\sigma_{eff}$ is defined by $$ \sigma_{eff} =
\sum_{i,j} \sigma_{ij} r_i r_j $$ and $\sigma_{ij}$ is the pair
annihilation cross section of the particle $\chi_i$ and $\chi_j$.
Here $r_i$ is given by $$ r_i = \frac{g_i (1+\Delta_i)^{3/2} e^{-\Delta_i
x}}{g_{eff}}$$
   
Due to the fact that the LSP is almost pure bino, the co-annihilation
processes go predominantly into fermions. However, since the coupling of
$\tilde{\chi}_2^0-f- \tilde{f}$ is proportional to
$Z_{2j}$, this coupling is smaller than the corresponding one of
$\tilde{\chi}_1^+-f- \tilde{f'}$. We found that  
the dominant contribution is due to the co-annihilation channel
$\tilde{\chi}_1^+
\chi \rightarrow f \bar{f}$. We also include the $\tilde{\chi}_1^+ \chi
\rightarrow W^+ \gamma$ channel which is estimated to give a few
percent contribution.
Then, we can  calculate the relic abundance from the equation
\begin{equation}
\Omega_{\chi} h^2 \simeq \frac{1.07 \times 10^{9}
\mathrm{GeV}^{-1}}{g_*^{1/2} M_P
x_F^{-1} \int_{x_f}^{\infty} \langle \sigma_{eff} v \rangle x^{-2} d x }.
\label{relic}
\end{equation}
Here $M_P$ is the Planck scale, $g_* \simeq 81$ is the effective number
of massless degrees of
freedom at freeze out and $x_F$ is given by
\begin{equation}
x_F = \ln \frac{0.038\ g_{eff}\ M_P\ (c+2) c\ m_{\tilde{\chi}_1^0}\ \langle \sigma_{eff} v
\rangle (x_F)}{g_*^{1/2} x_F^{1/2}},
\end{equation}
the constant $c$ is equal to $1/2$. In Figure(4) we plot the values of the
LSP relic abundance
$\Omega_{\chi} h^2$ values versus $m_{\chi}$. These values have been
estimated using eq.(\ref{relic}) with including the co-annihilations. 
\begin{figure}[h]
\psfig{figure=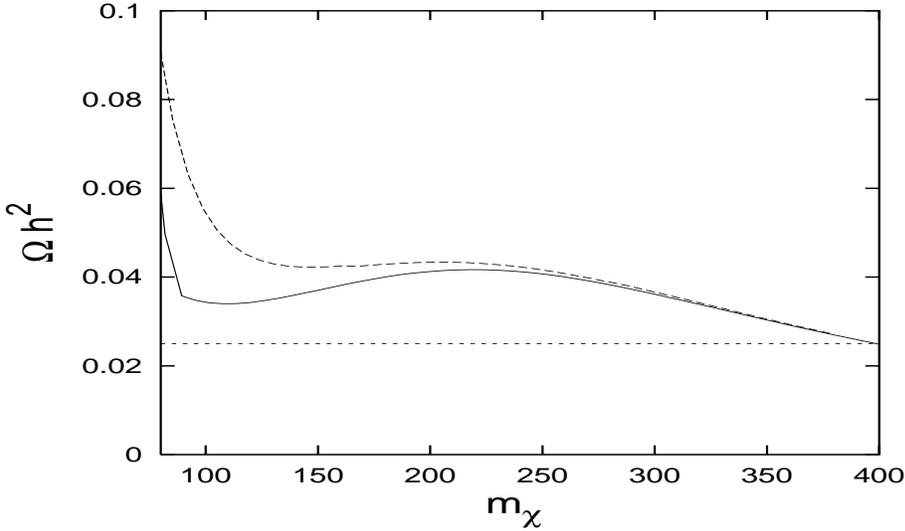,height=7cm,width=12cm}
\caption{The LSP relic abundance with co-annihilation versus its mass,
solid line corresponds
to non vanishing phases while the dashed lines correspond to vanishing phases. }
\end{figure}

The results in Figure (4)  show how the co-annihilation processes can play 
a crucial  rule in reducing the
values of $\Omega_{\chi} h^2$. By means of the lower bound on the relic
density  $\Omega_{\chi} h^2 > 0.025$  leads to  $m_{\chi} < 400$ GeV.
Also, here the effect of the SUSY phases is
insignificant and the same upper bound of the LSP mass is obtained for
vanishing and non vanishing phases.
\vskip 0.25cm
It is worth mentioning that the gaugino phases, especially the phase of
$M_3$, have a relevant impact on generating large $\phi_A$ at the
electroweak (EW) scale. It dominantly
contributes to the phase of the $A$-term during the renormalization
from the GUT scale to EW scale. Thus the radiative corrections to $\phi_A$
are very small and the phase of $A$ can be kept large at EW. However, as
we were shown, such large phases are not effecting for the LSP mass and
the relic abundance. In fact, this result can be explained as follow:  
the LSP is bino like (so it slightly depends on the phase of
$\mu$) and the contribution of the phases can be relevant if there is a
significant mixing in the sfermion mass
matrix. In the class of models we consider the off diagonal element are
much smaller than the diagonal element. 
\vskip 0.25cm
As shown in Ref.~\cite{shafi}, the SUSY phases have a significant effect
on the direct detection rate ($R$) and indirect detection rate ($\Gamma$):
the phase of $\phi_A$ increases the values of $R$ and $\Gamma$ .
Furthermore the enhancement of the ratios of the rates with non vanishing
$\phi_A$ to the  rates in the absence of this phase are even larger
than what is found in Ref.~\cite{shafi}. Indeed due to the gluino
contribution (through the renormalization) the phase $\phi_A$
can get larger values at EW scale.
\section{{\bf Conclusions}}
We considered type I string derived model which leads to non-universal
gaugino masses and phases. As recently shown, these non-universality is
very important to have sufficient
cancellations among different contributions to the EDM. Moreover the EDM
of the electron and neutron imposed constraint on the ratio of the gaugino
masses $M_1$ and $M_2$. This implies that the mass of the LSP (bino-like)
is close to the lightest chargino mass. The co-annihilation between the LSP
and lightest chargino is crucial to reduce the relic density to an
interesting region. The phases have no significant effect on the LSP mass
and its relic density, but have a substantial effect on the direct and
indirect detection rates. 

\section*{{\bf Acknowledgments}}
This work is supported by the Spanish Ministerio de Educacion y
Cultura research grant.

\providecommand{\bysame}{\leavevmode\hbox
to3em{\hrulefill}\thinspace}


\newpage

\begin{thebibliography}{99}

\bibitem{dugan}
M.~Dugan, B.~Grinstein, and L.~Hall, \npb{255}{85}{413}.
\vskip 0.2cm
\bibitem{nath} 
T.~Ibrahim and P.~Nath, \plb{418}{98}{98};\\
\prd{D57}{98}{478}; Erratum idid \textbf{D58} (1998) 019901.
\vskip 0.2cm  
\bibitem{barr}
S.~Barr and S.~Khalil {\it Phys. Rev.} {\bf D 61} (2000) 035005.
\vskip 0.2cm
\bibitem{kane}
M.~Brhlik, L.~Everett, G.~Kane, and J.~Lykken, \prl{83}{99}{2124} ;\\
M.~Brhlik, L.~Everett, G.~Kane, and J.~Lykken, hep-ph/9908326. 
\vskip 0.2cm
\bibitem{arnowitt}
E.~Accomando, R.~Arnowitt, and B.~Dutta , hep-ph/9909333. 
\vskip 0.2cm
\bibitem{shafi}
S.~Khalil and Q.~Shafi, \npb{564}{99}{19} .
\vskip 0.2cm
\bibitem{falk}
U.~Chattopadhyay, T.~Ibrahim, and P.~Nath, \prd{60}{99}{063505};\\ 
T.~Falk, A.~Frestl, and K.~Olive, \prd{59}{99}{055009} ;\\
T.~Falk and K.~Olive, \plb{375}{96}{196};\\
T.~Falk and K.~Olive, \plb{354}{95}{99} .
\vskip 0.2cm
\bibitem{munoz}
L.~Ibanez, C.~Munoz, and S.~Rigolin, \npb{553}{99}{43}. 
\vskip 0.2cm
\bibitem{randall}
J.~Feng, T.~Moroi, and L.~Randall, \prl{83}{99}{1731} ;\\
K.~Huitu, Y.~Kawamura, T.~Kobayashi, and K.~Puolamaki, hep-ph/9903528 ;\\
G.~Anderson, H.~Baer, C.~Chen, and X.~Tata,  hep-ph/9903370 ;\\
C.~Chen, M.~Drees, and J.~Gunion, \prd{55}{97}{347}, Erratum idid
\textbf{D60} (1999) 039901.
\vskip 0.2cm
\bibitem{george}
A.~Lahanas, D.~Nanopoulos, and V.~Spanos, hep-ph/9909497;\\ 
M.~Gomez, G.~Lazarides, and C.~ Pallis, hep-ph/9907261 ;\\
J.~Ellis, T.~Falk, K.~Olive, and M.~Srednicki, hep-ph/9905481.
\vskip 0.2cm
\bibitem{grist}
K.~Griest and D.~Seckel, \prd{43}{91}{3191}.
\vskip 0.2cm
\bibitem{report}
G.~Jungman, M.~Kamionkowski, and K.~Griest, \prep{267}{96}{195}.

\end{thebibliography}
\end{document}